\documentclass{amsart}

\usepackage{bm,bbm}
\usepackage{graphicx}

\def\ket#1{| #1 \rangle}

\def\kb#1#2{|#1\rangle\!\langle #2 |}
\def\cal#1{\mathcal{#1}}
\newcommand{\pfQED}{\rule{5pt}{5pt}}

\theoremstyle{plain}
\newtheorem{theorem}{Theorem}

\newtheorem{proposition}[theorem]{Proposition}
\newtheorem{lemma}[theorem]{Lemma}

\theoremstyle{definition}
\newtheorem{definition}[theorem]{Definition}
\newtheorem{example}[theorem]{Example}

\begin{document}

\title{Entropy of a quantum error correction code}

\author[David W. Kribs]{David W. Kribs$^{1}$}
\address{$^{1}$Department of Mathematics and Statistics, University of Guelph, Guelph, Ontario, N1G 2W1, Canada\\and Institute for Quantum Computing, University of Waterloo, Waterloo, Ontario, N2L 3G1, Canada}
\author[Aron Pasieka]{Aron Pasieka$^{2}$}
\address{$^{2}$Department of Physics, University of Guelph, Guelph, Ontario, N1G 2W1, Canada}
\author[Karol {\.Z}yczkowski]{Karol {\.Z}yczkowski$^{3}$}
\address{$^{3}$Instytut Fizyki im. Smoluchowskiego, Uniwersytet Jagiello{\'n}ski, ul. Reymonta 4, 30-059 Krak{\'o}w, Poland\\and Centrum Fizyki Teoretycznej, Polska Akademia Nauk, Al. Lotnik{\'o}w 32/44, 02-668 Warszawa, Poland}

\begin{abstract}
We define and investigate a notion of entropy for quantum error
correcting codes. The entropy of a code for a given quantum
channel has a number of equivalent realisations, such as through
the coefficients associated with the Knill-Laflamme conditions and
the entropy exchange computed with respect to any initial state
supported on the code. In general the entropy of a code can be
viewed as a measure of how close it is to the minimal entropy
case, which is given by unitarily correctable codes (including
decoherence-free subspaces), or the maximal entropy case, which from dynamical Choi matrix considerations corresponds to non-degenerate codes. We consider several examples, including a detailed analysis in the
case of binary unitary channels, and we discuss an extension of
the entropy to operator quantum error correcting subsystem codes.
\end{abstract}

\maketitle

\section{Introduction}

Quantum error correcting codes are a central weapon in the battle
to overcome the effects of environmental noise associated with
attempts to control quantum mechanical systems as they evolve in
time \cite{NC00,Got06}. It is thus important to develop techniques
that assist in determining whether one code is better than another
for a given noise model. In this paper we make a contribution to
this study by introducing a notion of entropy for quantum error
correcting codes.

No single quantity can be expected to hold all information on a
code, its entropy included. Nevertheless, the entropy of a code is
one way in which the amount of effort required to recover a code
can be quantified. In the extremal case, a code has zero entropy
if and only if it can be recovered with a single unitary
operation. This is the simplest of all correction operations in that a measurement
is not required as part of the correction process. These codes have been recently
coined {\it unitarily correctable}
\cite{KLPL05,KS06,BNPV07}, and include {\it decoherence-free
subspaces} \cite{DG97,KLV00a,LCW98a,Zan01b,ZR97c} in the case that recovery 
is the trivial identity operation. Thus more generally
the entropy can be regarded as a measure of how close a code is to
being unitarily correctable, or decoherence-free in some cases.

In the next section we introduce the entropy of a code, along with
required nomenclature. We also consider an example motivated by the stabilizer 
formalism \cite{Got96} and discuss an extension of code entropy to operator quantum
error correcting subsystem codes \cite{KLP05,Pou05,Bac05,KlaSar06}. 
We then consider in detail an illustrative
class of quantum operations for which the code structure has
recently been characterised, the class of {\it binary unitary
channels} \cite{CKZ06a,CKZ06b,CHKZ08,Woe07,LS08,CGHK08,Kazakov08}. 

\section{Entropy of a Quantum Error Correction Code}

Let $\rho$ denote a quantum state:
 a Hermitian, positive operator,   satisfying the trace
normalisation condition  Tr$\rho=1$. A linear quantum operation (or channel) 
$\Phi$, which sends a density operator $\rho$ of size $N$ into its
image $\rho'$ of the same size may be described in the Choi-Kraus
form \cite{Cho75a,Kr71}
\begin{equation}
\rho'=\Phi(\rho) =\sum_{i=1}^{M} E_i\rho E_i^{\dagger} .
 \label{Kraus}
\end{equation}
The Kraus operators $E_i$ can be chosen to be orthogonal
 $\langle E_i|E_j\rangle={\rm Tr}E_i^{\dagger}E_j
  = d_i \delta_{ij}$,
so that the non-negative weights $d_i$ become eigenvalues of the
dynamical (Choi) matrix, $D_\Phi=(\langle E_i|E_j\rangle )$. We
refer to the rank of the Choi matrix as the {\it Choi rank} of
$\Phi$. Observe that the Choi rank of $\Phi$ is equal to the minimal number 
of Kraus operators required to describe $\Phi$ as in (\ref{Kraus}).

Hence in the canonical form the number $M$ of non-zero Kraus operators does
not exceed $N^2$. Due to the theorem of Choi the map $\Phi$ is
completely positive (CP) if and only if the corresponding
dynamical matrix is positive if and only if $\Phi$ has a form as
in (\ref{Kraus}). The map $\Phi$ is {\it trace preserving}, ${\rm
Tr} \rho'={\rm Tr}\rho=1$, if and only if $\sum_{i=1}^{N^2}
E_i^{\dagger} E_i =\mathbbm{1}$ where we assume some $E_{i}$ are zero if $M$ is less than $N^{2}$. The family of operators $E_i$ is not
unique. However, if $\{F_j\}$ is another family of operators that
determine $\Phi$ as in (\ref{Kraus}), then there is a scalar
unitary matrix $U=(u_{ij})$ such that $E_i = \sum_j u_{ij} F_j$
for all $i$. We refer to this as the {\it unitary invariance} of
Choi-Kraus decompositions.

\subsection{Entropy Exchange and Lindblad Theorem}

To characterise the information missing in a quantum state one
uses its {\sl von Neumann} entropy, 
\begin{equation}\label{vNentropy}S(\rho) \ = \ - {\rm Tr} \rho
\log \rho.\end{equation} We will use the convention that $\log$ refers to logarithm base two as this provides a cleaner operational qubit definition in the context of quantum information.

In order to describe the action of a CP map $\Phi$,
represented in the canonical Choi-Kraus form (\ref{Kraus}), for an
initial state $\rho$ we may compare its entropy with the entropy
of the image $S(\rho')=S(\Phi(\rho))$. To obtain a bound for such
an entropy change we can define an operator
$\sigma=\sigma(\Phi,\rho)$ acting on an extended Hilbert space
${\cal H}_{N^2}$,
\begin{equation}
\sigma_{ij} \ = \ \mbox{Tr} \rho E_i^{\dagger} E_j , \quad
i,j=1,\dots, N^2 \ .
 \label{osigma}
\end{equation}
\noindent If the map $\Phi$ is stochastic, the operator $\sigma$
is positive definite and normalised so it represents a density
operator in its own right, $\sigma \in {\cal M}_{N^2}$ (specifically, it is an initially pure environment evolved by the unitary dilation of $\Phi$).  Observe
that for any unitary map, $\Phi_U(\rho)=U\rho U^{\dagger}$, the
form (\ref{Kraus}) consists of a single term only. Hence in this
case the operator $\sigma$ reduces to a single number equal to
unity and
 its entropy vanishes, $S\bigl(\sigma(\Phi_U,\rho)\bigr)=0$.

The auxiliary state $\sigma$ acting in an extended Hilbert space
was used by Lindblad to derive bounds for the entropy of an image
$\rho'=\Phi(\rho)$ of any initial state. The bounds of Lindblad
\cite{Li91},
\begin{equation}
0 \ \le \ |S(\rho')-S(\sigma)| \ \le \  S(\rho) \ \le \
S(\sigma)+S(\rho') \ ,
 \label{boundent}
\end{equation}
are obtained by defining another density matrix in the composite
Hilbert space ${\cal H}_N \otimes {\cal H}_{M}$,
\begin{equation}
\omega \: = \: \sum_{i=1}^{M} \sum_{j=1}^{M}
 E_j  \rho E_i^{\dagger} \otimes |i\rangle\langle j| \ ,
 \label{omega}
\end{equation}
\noindent where $M=N^2$ and $|i\rangle$ forms an orthonormal basis
in ${\cal H}_{M}$. Thus, $\omega$ is simply the system and an initially pure environment evolved by the unitary dilation of $\Phi$.  Computing partial traces one finds that Tr$_N
\omega=\sigma$ and Tr$_M \omega=\rho'$. It is possible to verify
that $S(\omega) = S(\rho)$, so making use of the subadditivity of
entropy and the triangle inequality  \cite{AL70} one arrives at
(\ref{boundent}).

If the initial state is pure, that is if $S(\rho) = 0$, we find
from (\ref{boundent}) that the final state $\rho'$ has entropy
$S(\sigma)$. For this reason $S(\sigma)$ was called the {\it
entropy exchange} of the operation by Shumacher \cite{Sc96}. In
that work an alternative formula for the entropy exchange was
proven,
\begin{equation}
S\bigl(\sigma(\Phi,\rho)) \ = \ S \Bigl(  ( \Phi \otimes {\mathbbm 1}) |\psi\rangle \langle \psi| \Bigr) \ ,
 \label{osigma2}
\end{equation}
\noindent where $|\psi\rangle$ is an \emph{arbitrary} purification
of the mixed state, Tr$_B |\psi\rangle \langle \psi|=\rho$.  Thus,
the entropy exchange is invariant under purification of the
initial state and remains a function only of the initial density
operator $\rho$ and the map $\Phi$.

\subsection{Quantum Error Correcting Codes}

A quantum operation $\Phi$ allows for an error correction scheme
in the standard framework for quantum error correction
\cite{Got96,Sho95a,Ste96a,BDSW96a,KL97} if there exists a subspace
$\cal C$ such that for some set of complex scalars $\Lambda =
(\lambda_{ij})$ the corresponding projection operator
$P_{\cal{C}}$ satisfies
\begin{eqnarray}
P_{\cal{C}} E_i^{\dagger} E_j P_{\cal{C}} \  = \  \lambda_{ij}
P_{\cal{C}} {\rm \quad  for  \quad all \quad} i,j=1,\dots,N^2.
\label{compression}
\end{eqnarray}
Specifically, this is equivalent to the existence of a quantum
{\sl recovery operation} $\Psi$ such that
\begin{equation}
\Psi \circ \Phi \circ {\cal{P}}_{\cal{C}} = {\cal{P}}_{\cal{C}},
\label{superoperator}
\end{equation}
where ${\cal{P}}_{\cal{C}}$ is the map ${\cal{P}}_{\cal{C}}(\rho)
= P_{\cal{C}} \rho P_{\cal{C}}$.  The subspace related to $P_{\cal{C}}$
determines a {\sl quantum error correcting code} (QECC) for the map
$\Phi$. A special class of codes are the {\it unitarily
correctable codes} (UCC), which are characterised by the existence
of a unitary recovery operation $\Psi(\rho) = U \rho U^\dagger$.
These codes include {\it decoherence-free subspaces} (DFS) in the
case that $\Psi$ is the identity map, $\Psi(\rho)=\rho$.

It can be shown that the matrix $\Lambda = (\lambda_{ij})$
is Hermitian and positive, and is in fact a density matrix, so it
can be considered as an auxiliary state acting in an extended
Hilbert space of size at most $N^2$. It is easy to obtain a more refined 
global upper bound on the rank of $\Lambda$ in terms of the map $\Phi$.

\begin{lemma}\label{upperbound}
Let $\Lambda$ be the matrix determined by a code $\mathcal C$ for
a quantum map $\Phi$. Then the rank of $\Lambda$ is bounded
above by the Choi rank of $\Phi$.
\end{lemma}

\noindent\textbf{Proof.} Without loss of generality assume the Choi matrix $D_\Phi$ is
diagonal. We have for all $i$,
\begin{equation}
\lambda_{ii} \dim {\mathcal C} = \mbox{Tr}(\lambda_{ii}P_{\mathcal{C}})= \mbox{Tr} (P_{{\mathcal C}}
E_i^\dagger E_i P_{{\mathcal C}}) \leq \mbox{Tr} (E_i^\dagger E_i)
= \langle E_i|E_i\rangle ,
\end{equation}
and the result follows from the positivity of $D_\Phi$ and
$\Lambda$.
\pfQED

Unitarily correctable codes are typically highly {\it degenerate} codes, as
the map $\Phi$ collapses to a single unitary operation when
restricted to the code subspace. In particular, the unitary
invariance of Choi-Kraus representations implies the restricted
operators $E_i P_{\mathcal C}$ are all scalar multiples of a
single unitary. More generally, one can see that a code ${\mathcal
C}$ is degenerate for  $\Phi$ precisely when the Choi rank of
$\Phi$ is strictly greater than the rank of $\Lambda$. Indeed, the
Choi rank counts the minimal number of Kraus operators required to
implement $\Phi$ via (2.1), and satisfaction of this strict
inequality means there is redundancy in the description of $\Phi
\circ {\mathcal P}_{\mathcal C} $ by the operators $E_i
P_{\mathcal C}$. Thus, for these reasons we shall refer to codes as {\it non-degenerate} if the Choi rank of $\Phi$ coincides with the rank
of $\Lambda$ and if the spectrum of $\Lambda$ is equally balanced -- that is to say that non-degenerate codes correspond to the maximally degenerate error correction matrix, $\Lambda$
proportional to the identity matrix

\subsection{Entropy of a Code}

Assume now that an error correcting code $\cal C$ exists and all
conditions (\ref{compression}) are satisfied. If a quantum state
$\rho$ is supported on the code then $P_{\cal{C}} \rho P_{\cal{C}}
= \rho$ and calculation of the entropy exchange (\ref{osigma})
simplifies,
\begin{equation}
\sigma_{ij} = \ \mbox{Tr} \rho E_i^{\dagger} E_j = \mbox{Tr}
P_{\cal{C}} \rho P_{\cal{C}}  E_i^{\dagger} E_j = \mbox{Tr} \rho
P_{\cal{C}}  E_i^{\dagger} E_j  P_{\cal{C}} = \mbox{Tr} \rho
\lambda_{ij} P_{\cal{C}}=\lambda_{ij} .
 \label{osigma3}
\end{equation}
In this way we have shown that the error correction matrix
$\Lambda$ is {\sl equal} to the auxiliary matrix $\sigma$ of
Lindblad, provided the initial state belongs to the code subspace.

From another direction, given an error correcting code $\cal{C}$
for a map $\Phi$, in \cite{KS06,NaySen06} it was shown that there
is a quantum state $\tau$ and an isometry $V$ such that for all
$\rho = P_{\cal{C}} \rho P_{\cal{C}}$,
\begin{equation}
\Phi(\rho) = V (\tau \otimes \rho) V^\dagger. \label{ssprinc}
\end{equation}
The result, which can be seen as a consequence of the decoupling condition of \cite{SW02}, gives an explicit way to ``see'' a code at the
output stage of a quantum process for which the code is
correctable. The result (and its subsystem generalization -- see
below) may also be viewed as a formalisation of the {\it subsystem
principle} for preserving quantum information \cite{KLV00a}. From
the proof of this result one can see directly that the entropy of
$\tau$ satisfies $S(\tau) = S(\Lambda)$.  This equality follows also from the fact that $\tau$ and $\Lambda$ can be interpreted as the states obtained
  by partial trace of an initially pure state with respect to two
  different subsystems.

Thus, from multiple perspectives we find motivation for the
following:

\begin{definition}\label{entdefn} Given a quantum operation $\Phi$ with Kraus operators
$\left\{E_{i}\right\}$ and a code $\cal{C}$ with matrix $\Lambda$
given by (\ref{compression}), we call the von Neumann entropy
$S(\Phi,{\cal{C}}):= S(\Lambda)$ the \emph{entropy of} ${\cal{C}}$
\emph{relative to} $\Phi$.
\end{definition}

The entropy of a code depends only on the map and the subspace
defined by $P_{\mathcal{C}}$, not on any particular state in the code
subspace.  Thus, the entropy exchange will be the same for all
initial states supported on the code subspace and is therefore a
property of the code itself.

In the following result we determine
what possible values the code entropy can take, and we derive a
characterisation of the extremal cases in terms of both the code
and the map.

\begin{theorem}\label{ucccase}
Let $\Phi$ be a quantum operation and let ${\mathcal C}$ be a code
with matrix $\Lambda$ given by (\ref{compression}). Then
$S(\Phi,{\cal{C}})$ belongs to the closed interval $[0,\log D]$,
where $D$ is the Choi rank of $\Phi$. Furthermore, the extremal
cases are characterised as follows:
\begin{itemize}
\item[$(i)$] $S(\Phi,{\cal{C}})=0$ if and only if ${\mathcal C}$
is a unitarily correctable code for $\Phi$.
\item[$(ii)$] $S(\Phi,{\cal{C}})=\log D$ if and only if $\mathcal
C$ is a non-degenerate code for $\Phi$.
\end{itemize}
\end{theorem}

\noindent\textbf{Proof.}
By Lemma~\ref{upperbound} and the subsequent discussion, the
maximal entropy case occurs when the rank of $\Lambda$ and
$D_\Phi$ coincide and the spectrum of $\Lambda$ is equally
balanced; that is, the code is non-degenerate. This occurs (by a standard spectral majorization
argument) precisely when the code entropy satisfies
$S(\Phi,{\cal{C}})=\log D$.

For the minimal entropy case, first suppose that $\cal{C}$ is a
UCC for $\Phi$. Then by (\ref{superoperator}) there is a unitary
operation ${\cal{U}}(\rho)= U\rho U^\dagger$ such that $\Phi \circ
{\cal{P}}_{\cal{C}} = {\cal{U}}\circ {\cal{P}}_{\cal{C}}$. Thus by
the unitary invariance of Choi-Kraus decompositions, it follows
that $E_iP_{\cal{C}} = \alpha_i U P_{\cal{C}}$ for some scalars
$\alpha_i$. Hence we have $\Lambda = (\overline{\alpha}_i
\alpha_j) = \kb{\psi}{\psi}$, where $\ket{\psi}$ is the
vector state with coordinates $\alpha_i$, and so
$S(\Phi,{\cal{C}})=S(\Lambda)=0$.

On the other hand, suppose $\Lambda = \kb{\psi}{\psi}$ is rank
one. Let $V$ be a scalar unitary that diagonalises $\Lambda$. It follows that 
$V$ induces a unitary change of representation for $\Phi$ via
(\ref{compression}) from $\{E_i\}$ to $\{F_j\}$. But since
$V\Lambda V^\dagger$ is diagonal, only one $F_j$,
say $F$, is non-zero, and hence by unitary invariance we have
$FP_{\cal{C}}=UP_{\cal{C}}$ for some unitary $U$. Thus $\Phi \circ
{\cal{P}}_{\cal{C}} = \cal{U}\circ {\cal{P}}_{\cal{C}}$, and the
result follows.
\pfQED

From an operational perspective, the numerical value of the entropy 
of a code allows us to quantify the number of ancilla qubits needed to perform a recovery operation.  
Specifically, the Choi rank, $D$, of a map $\Phi$ gives the minimum number of Kraus operators necessary 
to describe the map or, by Stinespring's dilation theorem~\cite{stinespring}, the dimension of the 
ancilla required to implement $\Phi$ as a unitary.  The rank of $\Lambda$ then gives the number of 
Kraus operators necessary to describe the action of the map restricted to $\mathcal{C}$ and thus the 
number of Kraus operators, $M$, necessary for a recovery operation in the usual measurement cum reversal
picture of recovery.  Again by Stinespring's dilation 
theorem we need an $M$-dimensional ancilla to implement the recovery as a unitary, and thus this requires $\log{M}$ qubits.

Hence the entropy is equal to zero for a unitarily correctable
code, in which the action of the
noise is unitary and thus requires no ancilla to implement the recovery operation. 
If the code entropy is positive, then any state of the code can potentially evolve to any
one of multiple locations in the system Hilbert space under the
action of the noise $\Phi$. This fact has to be compensated by the
recovery operation $\Psi$. The maximal entropy case for a particular $\Phi$ is
characterised by evolution to each of these locations ($M=D$ by Lemma~\ref{upperbound}) with equal
probability (by Theorem~\ref{ucccase}).  Here the entropy and thus the number of qubits in the ancilla will be $\log{D}$.

\subsection{Stabilizer Example}
As an example from the stabilizer
formalism, consider a three-qubit system with the usual notation
$X_i$, $Z_i$, $i=1,2,3$, for Pauli operators \cite{NC00}. The
single-qubit stabilizer code with generators
$\{Z_{1}Z_{2},Z_{2}Z_{3}\}$, is spanned by $|0_L\rangle = |000\rangle$ and
$|1_L\rangle = |111\rangle$. The set of operators
$\{\mathbbm{1},X_{1},X_{2},X_{3}\}$ form a correctable set of errors for
this stabilizer thus we can consider a channel, a three-qubit
bit-flip channel, comprised of these operators -- for example, the
channel with Kraus operators
$E_{1}=\sqrt{\frac{1}{3}(3-p-q-r)}\mathbbm{1}$,
$E_{2}=\sqrt{\frac{1}{3}p}X_{1}$, $E_{3}=\sqrt{\frac{1}{3}q}X_{2}$
and $E_{4}=\sqrt{\frac{1}{3}r}X_{3}$. Using
$P_{\mathcal{C}}=|000\rangle\langle 000| + |111\rangle\langle
111|$,~(\ref{compression}) tells us that the error correction
matrix is
$$\Lambda=\left(\begin{array}{cccc}\frac{1}{3}(3-p-q-r)&0&0&0\\0&\frac{1}{3}p&0&0\\0&0&\frac{1}{3}q&0\\0&0&0&\frac{1}{3}r\end{array}\right).$$
The entropy of this code is therefore
\begin{multline}S(\Lambda)=-\frac{1}{3}(3-p-q-r)\log{\frac{1}{3}(3-p-q-r)}
\\-\frac{1}{3}p\log{\frac{1}{3}p}-\frac{1}{3}q\log{\frac{1}{3}q}-\frac{1}{3}r\log{\frac{1}{3}r}.\notag\end{multline}
As we would expect, the minimum entropy is achieved when
$p=q=r=0$.

The maximum entropy occurs when \hbox{$p=q=r=\frac{3}{4}$},
which we might also expect since this puts the auxiliary density
matrix $\Lambda$ into the maximally mixed state.  Here, the Choi rank of our map is 
$4$ as is the rank of $\Lambda$, and the spectrum of $\Lambda$ is equally balanced. 
So $\mathcal{C}$ is non-degenerate for $\Phi$.  
Further, as we expect from Theorem~\ref{ucccase}, the entropy is \hbox{$\log{4}=2$}.  Since the 
rank of $\Lambda$ is the number of Kraus operators required for a recovery operation we 
find that in order to dilate the recovery operation to a unitary process, we need a 
$4$-dimensional space, or equivalently a $2$-qubit ancilla, which matches the value of the entropy.

We next consider a variant of this noise model, in which the third qubit undergoes no error and the first and second qubits are flipped with probability equal to that of no error. Thus the map $\Phi$ has noise operators $ \{ I, X_1, X_2 \}$, each weighted by $\frac{1}{\sqrt{3}}$. The entropies for a trio of correctable codes for $\Phi$ are given in Table~\ref{EntTable} (vectors are assumed to be normalised). The first code is non-degenerate, and thus yields the maximal entropy for this noise model. The second code has positive entropy less than the maximum since it is partially degenerate for $\Phi$. Indeed, one can check that $X_1$ (and $I$) act trivially on the code, whereas $X_2$ maps the code to an orthogonal subspace. The final code shows zero entropy since it is fully degenerate for $\Phi$. In fact, as can be directly verified it is a decoherence-free subspace for $\Phi$.

\begin{table}[h]

\caption{Codes for the noise model $\Phi = \frac{1}{\sqrt{3}} \{ I, X_1, X_2 \}$.}
\label{EntTable}
\begin{center}
\begin{tabular}{cc}

\hline

Qubit Code ${\mathcal C} = \{ |0_L \rangle , |1_L \rangle  \}$ & Entropy $S(\Phi,{\mathcal C})$ \\

\hline

$\{|000\rangle ,
|111\rangle\}$ & $\log 3$ \\

$\{|000\rangle + |100\rangle,
|011\rangle + |111\rangle\}$ & $\log 3 - \frac{2}{3}$ \\

$\{|000\rangle + |100\rangle + |010\rangle + |110\rangle,$ & $0$\\ 

$|011\rangle + |111\rangle + |001\rangle + |101\rangle \}$ & \\

\hline

\end{tabular}
\end{center}
\end{table}

\subsection{Entropy of a Subsystem Code}

We next consider an extension of code entropy to the case of
operator quantum error correcting subsystem codes. We shall only
introduce this notion here and leave a potential further investigation for
elsewhere. Subsystem codes were formally introduced under the
umbrella of {\it operator quantum error correction}
\cite{KLPL05,KLP05}, a framework that unifies the active and
passive approaches to quantum error correction. These codes now
play a central role in fault tolerant quantum computing.

Let $\cal{H}$ be a Hilbert space of finite dimension $N$.  Any decomposition $\cal{H} =
(A\otimes B)\oplus \cal{K}$ determines subsystems $A$ and $B$ of
$\cal{H}$. Given a quantum operation $\Phi$ acting on $\cal{H}$,
we say that a subsystem $B$ of $\cal{H}$ is {\it correctable} for
$\Phi$ if there exists maps $\Psi$ on $\cal{H}$ and $\tau_A$ on
$A$ such that
\begin{equation}\label{subsystemcode}
\Psi\circ \Phi \circ \cal{P}_{AB}= (\tau_A\otimes {\rm id}_B)
\circ \cal{P}_{AB},
\end{equation}
where ${\cal{P}}_{AB}(\rho)= P_{AB} \rho P_{AB}$ and $P_{AB}$ is
the projection onto the subspace $A\otimes B$.

Subsystem codes generalize standard (subspace) codes
(\ref{superoperator}) in the sense that subspaces may be regarded
as subsystems with trivial ancilla ($\dim A=1$). Thus, it is
natural to suggest that a notion of entropy for subsystem codes
should generalize the subspace definition. We find motivation for
such a notion through the main result of \cite{KS06} alluded to
above. To every correctable subsystem for $\Phi$, there are
subsystems $C$ and $B'\cong B$, a map $\tau_{C|A}$ from $A$ to $C$
and a unitary map $\cal{V}_{B'|B}$ from $B$ to $B'$ such that
\begin{equation}\label{KSresult}
\Phi \circ \cal{P}_{AB}= (\tau_{C|A}\otimes \cal{V}_{B'|B}) \circ
\cal{P}_{AB},
\end{equation}

Recall that the {\it maximal output entropy} of a channel $\Psi$ is
given by $S(\Psi) := \max_\rho S(\Psi(\rho))$. This motivates the
following.

\begin{definition}
Let $\Phi$ be a quantum operation with correctable subsystem $B$
that satisfies (\ref{subsystemcode}). Then we define the entropy
of $B$ relative to $\Phi$ as the maximal output entropy of the associated ancilla channel, 
$S(\tau_{C|A})$ from (\ref{KSresult}).
\end{definition}

Observe that this generalizes  Definition~\ref{entdefn}, since a
density operator may be regarded as a channel from a
one-dimensional input space. The minimal entropy case is
characterized by the ancilla channel $\tau_{C|A}$ having range
supported on a one-dimensional subspace. Such codes are unitarily
correctable, in fact as subspaces, but the converse is not true.
Instead, in the more general subsystem setting, the minimal
entropy case is described by the associated ancilla subsystem $A$
undergoing ``cooling'' to a fixed state. In principle one should
be able to conduct a deeper analysis of subsystem code entropy. We
leave this as an open investigation for elsewhere.

\section{Entropy of a Code for Binary Unitary Channels}

A binary unitary channel has the form \begin{equation}
\rho'=\Phi(\rho) = (1-p) W_1 \rho W_1^{\dagger} +
  p W_2 \rho W_2^{\dagger}
 \label{biunitary1}
\end{equation}
where $W_1$ and $W_2$ denote two arbitrary unitary operators and
the probability $p$  belongs to $[0,1]$.

It is clear that the problem of finding an error correcting code
subspace $\cal C$ for the above map is equivalent to the case
\begin{equation}
\rho'' = \Phi_U(\rho) = (1-p) \rho+
  p U \rho U^{\dagger}
 \label{biunitary2}
\end{equation}
where $U=W_1^{\dagger}W_2$. The number $M$ of Kraus operators
is equal to $2$, with $E_1=\sqrt{1-p} {\mathbbm 1}$ and
$E_2=\sqrt{p} U$. Thus the error correction  matrix $\Lambda$ is
of size two and reads
\begin{equation}
\Lambda \ = \ \left(\begin{matrix}
  1-p                                      &   \sqrt{p(1-p)} \lambda \\
 \sqrt{p(1-p)} \lambda^* &   p
\end{matrix}\right)
\label{lambda1}
\end{equation}
where $\lambda$ is a solution of the {\it compression problem} for
$U$
\begin{equation}
P_{\cal{C}}  U P_{\cal{C}}= \lambda P_{\cal{C}}  \ . \label{comp1}
\end{equation}

The set of solutions to this problem can be phrased in terms of
the {\it higher-rank numerical range} of the matrix $U$. The
rank-$k$ numerical range of $U$ is defined as
\begin{equation}
\label{rank-k-num-range}
\Omega_{k}(U)=\left\{\lambda\in\mathbb{C}\, |\, PUP=\lambda
P\text{ for some rank-$k$ projection } P \right\}.
\end{equation}
Given a dimension-$k$ that defines the size of the desired
correctable code, each $\lambda$ in $\Omega_{k}(U)$ corresponds to
a particular correctable code defined by the associated projection
$P$ that solves (\ref{comp1}). The following is a straightforward
application of (\ref{compression}).

\begin{proposition}
\label{correctable-if-numrange} Given a binary unitary channel
$\Phi$, there exists a rank-$k$ correctable code for $\Phi$ if and
only if the rank-$k$ numerical range of $U$ is non-empty.
\end{proposition}

Thus, the problem of finding the correctable codes for a given
binary unitary channel can be reduced to the problem of finding
the higher-rank numerical range of $U$.  This problem has recently
been solved in its entirety
\cite{CKZ06a,CKZ06b,CHKZ08,Woe07,LS08,CGHK08,Kazakov08}. Most succinctly, in terms
of the eigenvalues $\sigma(U)$ of $U$, the $k$th numerical range
of $U$ is the convex subset of the unit disk given by
\begin{equation}\label{omega1}
\Omega_k(U) \,\,\,= \,\,\,\bigcap_{\Gamma\,\subseteq\,
\sigma(U);\,\,|\Gamma|=N-k+1}\, {\rm conv}\, (\Gamma),
\end{equation}
where ${\rm conv}\,\{\lambda_1,\ldots,\lambda_m\}$ is the set of
linear combinations $\lambda = t_1\lambda_1 + \ldots
+t_m\lambda_m$ such that $\sum_{j=1}^m t_j = 1$ and   $t_j\geq 0$.
Figures~1.a and 1.b depict the case of a generic two-qubit unitary
($N=4$) with $k=2$, while Figure 1.c shows the case of a generic
two-qutrit unitary noise ($N=3\times 3 = 9$) with $k=3$.

\begin{figure}
\caption{Higher rank numerical range $\Omega_k$
for unitary matrices $U$ describing a bi-unitary channel: two
qubit system, a) example 2 with $\lambda=0$, b) case with
$r=|\lambda| >>0$
    for which code entropy is smaller,
c) two qutrit case with $\lambda\in \Omega_3(U)$ chosen
     to maximize its modulus $r$ and to minimize
   the code entropy $S(\Lambda)$.}\label{range08a}
\begin{center}
\includegraphics[width=325pt]{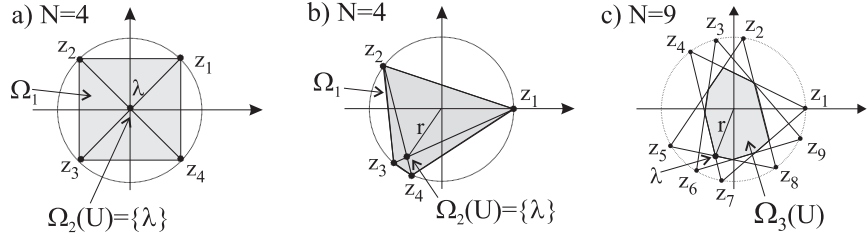}
\end{center}
\end{figure}

With a particular $\lambda$ in-hand, straightforward algebra
provides us with the spectrum of the matrix (\ref{lambda1}),
\begin{equation}
\Lambda_{\pm} \ = \  \frac{1}{2} \left(1 \pm  \sqrt{
1-4p(1-p)(1-|\lambda|^2)} \right),
 \label{lambda3}
\end{equation}
which allows us to calculate the entropy of the code,
\begin{equation}
S(\Phi,{\cal C}) = S(\Lambda) \ = \ - \Lambda_+ \log \Lambda_+
                           - \Lambda_- \log \Lambda_-.
 \label{lambda5}
\end{equation}
We are then led to the following:
\begin{theorem}
\label{entropy-biunitary} A minimum entropy rank-$k$ code for a
binary unitary channel $\Phi(\rho)=(1-p) \rho  +
  p U \rho U^{\dagger}$ corresponds to any
code for which the magnitude $|\lambda|$ of the compression values
$\lambda\in\Omega_k(U)$ is closest to unity, while the maximum
entropy corresponds to $|\lambda|$ closest to zero. Moreover, a
code with minimal entropy can be constructively obtained.
\end{theorem}
\noindent\textbf{Proof.}
The first statement follows directly from an application of first
and second derivative tests on (\ref{lambda5}), constrained to the
unit disk. A minimal entropy code can be explicitly constructed
based on the analysis of higher-rank numerical ranges
\cite{CKZ06a,CKZ06b,CHKZ08,Woe07,LS08,CGHK08,Kazakov08}.
\pfQED

\begin{example}
As an illustration of the code
construction in the simplest possible case ($N=4,k=2$), let $U$ be
the unitary with spectrum depicted in Figure~1.a, $\sigma(U) = \{
z_k = \exp (k\frac{\pi}{4}) : k=1,3,5,7\}$. Let $\ket{\psi_k}$ be
the associated eigenstates, $U\ket{\psi_k} = z_k\ket{\psi_k}$.  In
this case we have $\Omega_2(U)=\{0\}$ so $\lambda=0$, and one can
check directly that a single qubit correctable code for $\Phi$ is
given by ${\cal{C}} = {\rm span}\,\{ \ket{\phi_1}, \ket{\phi_2}
\}$, where
$$\left\{\begin{array}{lcl}\ket{\phi_1} &=& \frac{1}{\sqrt{2}}
(\ket{\psi_1}+\ket{\psi_3})\\
\ket{\phi_2} &=& \frac{1}{\sqrt{2}}
(\ket{\psi_2}+\ket{\psi_4})\end{array}.\right.$$ For a concrete
example, in the case that $p=0.01$, (\ref{lambda5}) yields
a code entropy of \hbox{$S(\Phi,{\cal{C}}) = 0.081$}.

The general case requires a more delicate construction,
nevertheless it can be done. The ``eigenstate grouping'' procedure
used above can be applied whenever $k$ divides $N$. For instance,
in the generic $N=9$ and $k=3$ case depicted in Figure~1.c, a
single qutrit code can be constructed for all $\lambda$ in the
region $\Omega_3(U)$. The states $\ket{\phi_i}$, $i=1,2,3$, can be
constructed in an analogous manner by grouping the nine
eigenstates for $U$ into three groups of three, and writing
$\lambda$ in three different ways as a linear combination of the
associated unimodular eigenvalues $z_j$, $j=1,\dots,9$.

However, without going into the details of this construction we
can still analyze the corresponding code entropies. For simplicity
assume the nine eigenvalues are distributed evenly around the unit
circle with $z_1=0$. By Theorem~\ref{entropy-biunitary} we know
that the entropy will be minimized for any $\lambda$ that gives
the minimum distance from $\Omega_3(U)$ to the unit circle. An elementary
calculation shows that one such $\lambda$, given by the
intersection of the lines through the first and seventh, and sixth
and ninth eigenvalues (counting counterclockwise) is approximately
$\lambda_0 = 0.092 - 0.524\,i$. With the probability $p=0.01$, the
corresponding error correction matrix $\Lambda$ has spectrum
$\{0.007,0.993\}$. Thus, (\ref{lambda3}) and (\ref{lambda5}) yield
the minimal qutrit code entropy for this channel as
$$\min_{\dim{\cal{C}}=3} S(\Phi,{\cal{C}})=
\min_{\lambda\in\Omega_3(U)} S(\Lambda) =
S(\Lambda_{\lambda=\lambda_0}) = 0.060.$$ On the other hand, as
$\lambda =0$ belongs to $\Omega_3(U)$, by
Theorem~\ref{entropy-biunitary} and (\ref{lambda1}) we also see
the maximal entropy for $p=0.01$ occurs for any code with
$\lambda=0$. In such cases we have $\Lambda$ spectrum $\{0.01,
0.99\}$, and hence the maximal entropy is $S(\Lambda_{\lambda=0})
= 0.081$.

Changing focus briefly, if we fix an arbitrary unitary $U$, then
we could consider the family of channels determined by varying the
probability $p$. It follows from (\ref{lambda5}) that the channel
with the correctable codes of maximal entropy corresponds to the
$p=\frac{1}{2}$ channel, and the channels whose correctable codes
possess minimal entropy correspond to $p=0$ and $p=1$. Indeed, the
value of $\lambda$ depends on $U$ but not $p$, thus a given
$\lambda$ will solve (\ref{comp1}) for any $p$ and so $\lambda$
can be chosen independently using the above theorem. The result
once again follows from application of first and second derivative
tests with $p$ between $0$ and $1$. 
\end{example}

The following results show that the entropy of a code for a binary unitary channel can be regarded as a measure of how close the code
is to a decoherence-free subspace.

\begin{lemma}\label{uccdfs}
If $\Phi$ is a binary unitary channel, then the sets of unitarily
correctable subspaces and decoherence-free subspaces coincide.
\end{lemma}
\noindent\textbf{Proof.}
As proved in \cite{KS06}, for a bistochastic (unital) map $\Phi$,
a condition satisfied by every binary unitary channel, the
unitarily correctable codes (respectively the decoherence-free
subspaces) for $\Phi$ are imbedded in the fixed point algebra for
the map $\Phi^\dagger\circ\Phi$ (respectively $\Phi$), where
$\Phi^\dagger$ is the Hilbert-Schmidt dual map of $\Phi$. In
particular, it follows from this fact that the former set is given
by the set of operators that commute with $U$ and $U^\dagger$,
whereas the latter is the set of operators that commute with $U$.
By the Spectral Theorem these two sets coincide.
\pfQED

\begin{theorem}\label{dfsbuc}
Let $\Phi$ be a binary unitary channel. Then there is a rank-$k$
code $\cal{C}$ of zero entropy, $S(\Phi,{\cal{C}})=0$, for $\Phi$
if and only if there is a $k$-dimensional decoherence-free
subspace for $\Phi$ if and only if there exists
$\lambda\in\Omega_k(U)\cap\sigma(U)$.
\end{theorem}
\noindent\textbf{Proof.}
A $k$-dimensional decoherence-free subspace for $\Phi$ corresponds
to an eigenvalue $\lambda$ of $U$ with multiplicity at least $k$
(see \cite{KS06} and the references therein); that is,
$\lambda\in\Omega_k(U)\cap\sigma(U)$. The rest follows from the
lemma and previous theorem.
\pfQED

In order to further illustrate these results, consider again the
case of an arbitrary two-qubit system ($N=4$).  The correctable
codes with largest entropy are those with $p=\frac{1}{2}$ and so
the spectrum of $\Lambda$ reads
\begin{equation}
\Lambda_{\pm} \ = \  \frac{1}{2} \bigl(1 \pm |\lambda| \bigr).
 \label{lambda4}
\end{equation}
In the two-qubit case, the complex number $\lambda$ is given by
the point inside the unit circle at which two diagonals of the
quadrangle formed by the spectrum of $U$ cross (see Figures~1.a and
1.b). Consider a special case of the problem where $U$ has a
doubly degenerated eigenvalue, so that $|\lambda|=1$. For example,
$U$ could be any (non-identity) element of the two-qubit Pauli
group. Then the spectrum of $\Lambda$ consists of $\{1,0\}$
which implies  $S(\Lambda)=0$ (despite $p$ having been chosen for
the largest entropy correctable codes). Hence $\Lambda$ is pure
and there exists a decoherence free subspace -- the one spanned by
the degenerated eigenvalues of $U$.

In general, for binary unitary channels one may use the entropy
(\ref{lambda5}) as a measure quantifying to what extend a given
error correction code is close to a decoherence-free subspace. For
instance, any channel (\ref{biunitary2}) acting on a two qubit
system and described by unitary matrix
 $U=W_1^{\dagger}W_2$ of size $4$
may be characterized by the radius $r=|\lambda|$ of the point in
which two diagonals of the quadrangle of the spectrum cross.
The larger $r$, the smaller entropy $S(\Phi,{\cal C})$, and the closer
the error correction code is to a decoherence-free space.

The code entropy can be also used to classify
codes designed for a binary unitary channel acting on larger
systems. For instance in the case of two qutrits,
 $N=3 \times  3=9$,
one can find a subspace supported on $k=3$ dimensional subspace.
The solution is by far not unique and can be parametrized by
complex numbers $\lambda$ belonging to an intersection of $3$
triangles, which forms a convex set of a positive measure. From
this set one can thus select a concrete solution  providing a code
$\cal C$, such that  $r=|\lambda|$ is the largest, which implies
that the code entropy,  $S(\Phi,{\cal C})$, is the smallest -- see Figure~1.c. Such an error correction code is distinguished by being as close to a decoherence-free
subspace as possible.

\section{Conclusions}

We have investigated a notion of entropy for quantum error
correcting codes and quantum operations. The entropy has multiple
natural realisations through fundamental results in the theory of
quantum error correction. We showed how the extremal cases are
characterised by unitarily correctable codes and decoherence-free
subspaces on the one hand, and the non-degenerate case
determined by the Choi matrix of the map on the other. We
considered examples from the stabilizer formalism, and conducted a
detailed analysis in the illustrative case of binary unitary
channels. Recently developed techniques on higher-rank numerical
ranges were used to give a complete geometrical description of
code entropies for binary unitary channels; in particular, the
structure of these subsets of the complex plane can be used to
visually determine how close a code is to a decoherence-free
subspace. We also introduced an extension of code entropy to
subsystem codes, and left a deeper investigation of this notion
for elsewhere. It could be interesting to explore further
applications of the code entropy in quantum error correction. For
instance, although quantum error correction codes were originally
designed for models of discrete time evolution in the form of a
quantum operation, generalizations to the
case of continuous evolution in time \cite{Bra98,LS98,OLB08} have been investigated.
Further, we have investigated perfect correction codes only, for
which the error recovery operation brings the quantum state
corrupted by the noise back to the initial state with fidelity
equal one. Such perfect correction codes may be treated as a
special case of more general approximate error
correction codes \cite{SW02,CGS05,Kl07}. Another recent investigation \cite{Ali08} includes 
analysis that suggests the measurement component of recovery may prove to be problematic 
in quantum error correction, and hence may motivate further investigation of unitarily correctable codes.

\section{Acknowledgements}

We thank the referee for helpful comments.  D.W.K. was partially supported by NSERC grant 400160, by NSERC
Discovery Accelerator Supplement 400233, and by Ontario Early Researcher
Award 48142. A.P. was partially supported by an Ontario Graduate
Scholarship. K.{\.Z}. acknowledges support of an European research
project SCALA and the special grant number  DFG-SFB/38/2007 of
Polish Ministry of Science.


\begin{thebibliography}{10}

\bibitem{NC00}
M.~A. Nielsen and I.~L. Chuang.
\newblock {\em Quantum Computation and Quantum Information}.
\newblock Cambridge, New York, 2000.

\bibitem{Got06}
D.~Gottesman.
\newblock Quantum error correction and fault tolerance.
\newblock In J.-P. Francoise, G.L. Naber, and S.T. Tsou, editors, {\em
  Encyclopedia of Mathematical Physics}, volume~4, page 196. Oxford, Elsevier,
  2006, quant-ph/0507174.

\bibitem{KLPL05}
D.~W. Kribs, R.~Laflamme, D.~Poulin, and M.~Lesosky.
\newblock Operator quantum error correction.
\newblock {\em Quantum Inf. Comput.}, 6(4\&5):382, 2006, quant-ph/0504189.

\bibitem{KS06}
D.~W. Kribs and R.~W. Spekkens.
\newblock Quantum error-correcting subsystems are unitarily recoverable
  subsystems.
\newblock {\em Phys. Rev. A}, 74:042329, 2006, quant-ph/0608045.

\bibitem{BNPV07}
R.~Blume-Kohout, H.K. Ng, D.~Poulin, and L.~Viola.
\newblock The structure of preserved information in quantum processes.
\newblock {\em Phys. Rev. Lett.}, 100:030501, 2008, 0705.4282.

\bibitem{DG97}
L.-M. Duan and G.-C. Guo.
\newblock Preserving coherence in quantum computation by pairing quantum bits.
\newblock {\em Phys. Rev. Lett.}, 79:1953, 1997.

\bibitem{KLV00a}
E.~Knill, R.~Laflamme, and L.~Viola.
\newblock Theory of quantum error correction for general noise.
\newblock {\em Phys. Rev. Lett.}, 84:2525, 2000.

\bibitem{LCW98a}
D.A. Lidar, I.L. Chuang, and K.B. Whaley.
\newblock Decoherence free subspaces for quantum computation.
\newblock {\em Phys. Rev. Lett.}, 81:2594, 1998, quant-ph/9807004.

\bibitem{Zan01b}
P.~Zanardi.
\newblock Stabilizing quantum information.
\newblock {\em Phys. Rev. A}, 63:12301, 2001, quant-ph/9910016.

\bibitem{ZR97c}
P.~Zanardi and M.~Rasetti.
\newblock Noiseless quantum codes.
\newblock {\em Phys. Rev. Lett.}, 79:3306, 1997, quant-ph/9705044.

\bibitem{Got96}
D.~Gottesman.
\newblock A class of quantum error-correcting codes saturating the quantum
  hamming bound.
\newblock {\em Phys. Rev. A}, 54:1862, 1996, quant-ph/9604038.

\bibitem{KLP05}
D.~Kribs, R.~Laflamme, and D.~Poulin.
\newblock A unified and generalized approach to quantum error correction.
\newblock {\em Phys. Rev. Lett.}, 94:180501, 2005, quant-ph/0412076.

\bibitem{Pou05}
D.~Poulin.
\newblock Stabilizer formalism for operator quantum error correction.
\newblock {\em Phys. Rev. Lett.}, 95:230504, 2005, quant-ph/0508131.

\bibitem{Bac05}
D.~Bacon.
\newblock Operator quantum error correcting subsystems for self-correcting
  quantum memories.
\newblock {\em Phys. Rev. A}, 73:012340, 2006, quant-ph/0506023.

\bibitem{KlaSar06}
S.A. Aly, A.~Klappenecker, and P.K. Sarvepalli.
\newblock Subsystem codes.
\newblock {\em prepint arXiv:quant-ph/0610153}, 2006.

\bibitem{CKZ06a}
M.-D. Choi, D.W. Kribs, and K.~{\.Z}yczkowski.
\newblock Higher-rank numerical ranges and compression problems.
\newblock {\em Linear Algebra Appl.}, 418:828--839, 2006, math/0511278.

\bibitem{CKZ06b}
M.-D. Choi, D.W. Kribs, and K.~{\.Z}yczkowski.
\newblock Quantum error correcting codes from the compression formalism.
\newblock {\em Rep. Math. Phys.}, 58:77--91, 2006, quant-ph/0511101.

\bibitem{CHKZ08}
M.-D. Choi, J.A. Holbrook, D.W. Kribs, and K.~{\.Z}yczkowski.
\newblock Higher-rank numerical ranges of unitary and normal matrices.
\newblock {\em Oper. Matrices}, 1:409--426, 2007, quant-ph/0608244.

\bibitem{Woe07}
H.~Woerdeman.
\newblock The higher rank numerical range is convex.
\newblock {\em Linear Multilinear Algebra}, 56:65--67, 2008.

\bibitem{LS08}
C.-K. Li and N.-S. Sze.
\newblock Canonical forms, higher rank numerical ranges, totally isotropic
  subspaces, and matrix equations.
\newblock {\em Proc. Amer. Math. Soc.}, 136:3013--3023, 2008, 0706.1536.

\bibitem{CGHK08}
M.-D. Choi, M.~Giesinger, J.A. Holbrook, and D.W. Kribs.
\newblock Geometry of higher-rank numerical ranges.
\newblock {\em Linear Multilinear Algebra}, 56:53--64, 2008.

\bibitem{Kazakov08}
A.Y. Kazakov.
\newblock Elementary constructive approach to the higher-rank numerical ranges
  of unitary matrices.
\newblock {\em J. Phys. A}, 41(25):255306, 2008, 0707.0170.

\bibitem{Cho75a}
M.-D. Choi.
\newblock Completely positive linear maps on complex matrices.
\newblock {\em Linear Algebra Appl.}, 10:285, 1975.

\bibitem{Kr71}
K.~Kraus.
\newblock General state changes in quantum theory.
\newblock {\em Ann. Phys.}, 64:311, 1971.

\bibitem{Li91}
G.~Lindblad.
\newblock Quantum entropy and quantum measurements.
\newblock In C.~Bendjaballah et~al., editor, {\em Lecture Notes in Physics},
  volume 378, page~36, Berlin, 1991. Springer-Verlag.

\bibitem{AL70}
H.~Araki and E.~Lieb.
\newblock Entropy inequalities.
\newblock {\em Comm. Math. Phys.}, 18(2):160, 1970.

\bibitem{Sc96}
B.~Schumacher.
\newblock Sending entanglement through noisy quantum channels.
\newblock {\em Phys. Rev. A}, 54:2614, 1996, quant-ph/9604023.

\bibitem{Sho95a}
P.~W. Shor.
\newblock Scheme for reducing decoherence in quantum computer memory.
\newblock {\em Phys. Rev. A}, 52:R2493, 1995.

\bibitem{Ste96a}
A.~M. Steane.
\newblock Error correcting codes in quantum theory.
\newblock {\em Phys. Rev. Lett.}, 77:793, 1996.

\bibitem{BDSW96a}
{C.~H.} Bennett, D.~P. DiVincenzo, J.~A. Smolin, and W.~K. Wootters.
\newblock Mixed state entanglement and quantum error correction.
\newblock {\em Phys. Rev. A}, 54:3824, 1996, quant-ph/9604024.

\bibitem{KL97}
E.~Knill and R.~Laflamme.
\newblock A theory of quantum error-correcting codes.
\newblock {\em Phys. Rev. A}, 55:900, 1997, quant-ph/9604034.

\bibitem{NaySen06}
A.~Nayak and P.~Sen.
\newblock Invertible quantum operations and perfect encryption of quantum
  states.
\newblock {\em Quantum Inf. Comput.}, 7(1\&2):103, 2007, quant-ph/0605041.

\bibitem{SW02}
B.~Schumacher and M.~D. Westmoreland.
\newblock Approximate quantum error correction.
\newblock {\em Quantum Inf. Process.}, 1(1-2):5, 2002, quant-ph/0112106.

\bibitem{stinespring}
W.F. Stinespring.
\newblock Positive functions on ${C}^{*}$-algebras.
\newblock {\em Proc. Amer. Math. Soc.}, 6:211, 1955.

\bibitem{Bra98}
S.L. Braunstein.
\newblock Error correction for continuous quantum variables.
\newblock {\em Phys. Rev. Lett.}, 80:4084, 1998, quant-ph/9711049.

\bibitem{LS98}
S.~Lloyd and J.-J.E. Slotine.
\newblock Analog quantum error correction.
\newblock {\em Phys. Rev. Lett.}, 80:4088, 1998, quant-ph/9711021.

\bibitem{OLB08}
O.~Oreshkov, D.~A. Lidar, and T.~A. Brun.
\newblock Operator quantum error correction for continuous dynamics.
\newblock {\em preprint arXiv:0806.3145}, 2008.

\bibitem{CGS05}
C.~Crepeau, D.~Gottesman, and A.~Smith.
\newblock Approximate quantum error-correcting codes and secret sharing
  schemes.
\newblock {\em preprint arXiv:quant-ph/0503139}, 2005.

\bibitem{Kl07}
R.~Klesse.
\newblock Approximate quantum error correction, random codes, and quantum
  channel capacity.
\newblock {\em Phys. Rev. A}, 75:062315, 2007, quant-ph/0701102.

\bibitem{Ali08}
R.~Alicki.
\newblock Quantum decay cannot be completely reversed. {T}he 5\% rule.
\newblock {\em preprint arXiv:0807.2609}, 2008.

\end{thebibliography}
\end{document}